\documentstyle[dcolumn]{l-aa} 

\def\min{\raisebox{0.4ex}{$\scriptscriptstyle{-}$}}
\def\plu{\raisebox{0.4ex}{$\scriptscriptstyle{+}$}}
\def\plumi{\raisebox{0.4ex}{$\scriptscriptstyle{\pm}$}}
\def\et{{et\thinspace al.}\ }                     
\def\kms{km\thinspace s$^{-1}$ }                  
\def\kmx{km\thinspace s$^{-1}$}                   
\def\amm{\AA\thinspace mm$^{-1}$ }                
\def\onerule{\noalign{\medskip\hrule\medskip}}        

\begin{document}

\thesaurus{03        
           (04.03.1; 
            11.03.1; 
            11.04.1) 
}

\title{Structure and kinematics of galaxy clusters}

\subtitle{I. The redshift catalogue
\thanks{Based on observations made at the European Southern Observatory,
La Silla, Chile.}}

\author{P.~Stein
\thanks{{\it Present address:\/}
Departament d'Astronomia i Meteorologia, Universitat de Barcelona,
Avenida Diagonal 647, E--08028 Barcelona, Spain.
E-mail: paul at faess0.am.ub.es}
}

\offprints{P.~Stein (Universitat de Barcelona)}

\institute{Astronomisches Institut der Universit\"at Basel, 
Venusstrasse 7, CH--4102 Binningen, Switzerland}

\date{Received ...; accepted ...}

\maketitle

\begin{abstract}

An extensive redshift survey has been conducted on a sample of
15 nearby ($0.01\la z\la 0.05$) clusters of galaxies.
A total number of 860 redshifts were determined 
by fitting of emission--lines and/or cross-correlation techniques.
Of this sample, 735 galaxies are within 
0.2--0.8 Mpc ($H_0$ = 50 \kmx Mpc$^{-1}$) of the center of clusters. 
Approximate morphological types are available for most of the
galaxies. A comparison of the present redshifts with published data allows 
an extensive error analysis. The agreement is excellent with the most
modern data, showing a zero point error of 5 \kms and an overall
consistency of the measurements and their uncertainties. We estimate our
redshifts to have mean random errors around 
30 \kmx. A population analysis of the clusters will be 
given in a forthcoming paper.\footnote{Tables 1 and 2 are only available in
electronic form at the CDS via anonymous ftp 130.79.128.5}


\end{abstract}

\input boxedeps
\EmulateRokicki
\SetepsfEPSFSpecial

\section{Introduction}
During the last few years the number of known redshifts has undergone an 
exponential rise. This is due mainly to the strongly rising number of 
multi-fibre instruments in operation, and to the breakthrough of CCD detectors 
in combination with cross-correlation methods for measuring absorption-line 
redshifts.\\
Several redshift surveys aimed at studying galaxy clusters, which led to
significant progress in the understanding of structural and dynamical 
aspects (e.g. Lucey \et 1986; Dressler \& Shectman 1988; 
Beers \et 1991b; Malumuth \et 1992). 
Most of these studies concentrated on measuring redshifts for 
the brightest galaxies in regions extending $\ga 1$ Mpc around the cluster
centers. Thus little is known 
about the degree of substructure in the cores of 
galaxy clusters, an information which would greatly constrain theories of 
dark matter distribution and cluster formation and evolution 
\mbox{(Merritt 1988)}.
Other challenging questions about galaxy clusters remain 
yet unanswered. Among them: what is
the kinematical status of galaxies of different 
types, especially as compared with the morphological segregation in projected
space
(Dressler \mbox{1980) ?} Which frequency and significance have
peculiar velocities of cD galaxies ? Is mass segregation 
in the very center of rich clusters a dominant effect ?\\
As a contribution towards the solution of these problems we have measured
redshifts for about 30--100 galaxies in the central regions of several
nearby galaxy clusters ($0.01\la z\la 0.05$). The nearness of our 
sample allows for raw, mainly new morphological types to be assigned 
using existing photographic material.

\section{Observations}
Selecting an unbiased and possibly complete sample of nearby galaxy clusters
was difficult when this project began in 1988. The 
southern Abell cluster catalogue (Abell \et 1989) 
was not yet published. Moreover, neither X-ray selected
cluster catalogues (e.g. Gioia \et 1990) nor catalogues based on 
automated selection criteria (e.g. Lumsden \et 1992) 
were available at that
time. Our cluster sample was chosen inspecting ESO survey plates by eye,
and selecting the highest galaxy density peaks on them.
Most of the clusters later turned out to
fulfill Abell's criteria. This sample is representative in the sense that
it contains a significant fraction of nearby clusters of different morphologies,
richnesses and X-ray luminosities.\\
All galaxy spectra were obtained with the OPTOPUS multifibre instrument.
The two observing runs on 
the \mbox{3.6 m} telescope at ESO, La Silla, took place in August, 1989 and 
April/May, 1990. 
The main characteristics of OPTOPUS are: a field of view of 
33\arcmin\ and as many as 30 (1989) respectively 50 (1990) fibres, each 
of 2.3\arcsec\ diameter. These fibres were plunged into holes that had
been previously drilled into a solid plate.
Two different sets of fibres were provided during the 1990 run, thus 
allowing to plunge one set of fibers into a drilled plate, while
the other fibre set was performing at the telescope. This 
significantly reduced the time loss between two subsequent exposures.
Excellent weather conditions allowed a total of 34 exposures in 8 
nights to be taken. During these observations 1276 spectra were 
obtained in 20 fields. 
Although some of the fields revealed themselves as chance 
projections of partly unrelated galaxies in the plane of the sky, the great 
majority of the acquired data proved to be inside rich clusters of galaxies.
\\
During the first observing run in 1989 a 
dispersion of 170 \amm (first order) was chosen, which together with a 
1024$\times$640 CCD and a pixel size of 15 $\mu$m gives a 
nominal resolution of 2.55 
\AA/pixel. The wavelength range was chosen between 3800 \AA\  and 5800 \AA.
In the 1990 run a dispersion of 224 \amm was preferred, which 
leads to a resolution of 3.36 \AA/pixel. 
A broader range in wavelength, between 3800 \AA\
 and 6500 \AA\ was thus covered. On both occasions the inclusion of 
some bright emission lines 
(H$_\delta$, H$_\gamma$, H$_\beta$, O~III) and most of 
the absorption features suitable for cross-correlation 
(Ca~II H\plu K, G-band, H$_\beta$, Mg~I, in 1990 also Na~D and several iron 
lines) was guaranteed. \\
For each field a single exposure with a duration of 70 min was taken. 
He--Ne wavelength calibrations as well as night-sky exposures preceded and 
followed each exposure.

\section{Data reduction}
A software package for automated processing of multifibre spectra and 
subsequent cross-correlation has been developed by the author 
for reducing the data.
The only necessary on-line interactions are the 
measurement of emission line redshifts and also 
cutting evident emission lines prior to 
cross-correlation. The package consists mainly of FORTRAN programs with the
addition of some 
\mbox{MIDAS} commands and runs inside the \mbox{MIDAS} environment. 
An important consequence of 
this automated processing is the homogeneity and reproducibility of the 
results. Using FORTRAN routines instead of \mbox{MIDAS} procedures 
for the main steps of the cross-correlation 
has the advantage of speeding up the reduction: on a $\mu$VAX 4000 the time 
required to obtain redshifts from raw data is around 1 min/galaxy, once
the setup is done.\\
Starting from raw CCD spectra several reduction steps were necessary, 
including cleaning from defects, 
one-dimensional extraction, flat-fielding, wavelength 
calibration and sky subtraction. Thereafter, the spectral data were ready for 
redshift measurements, 
which was done either using emission lines or absorption lines or 
both, where available. A detailed description follows:\\
a) After pre-reduction of the CCD image much care was spent on optimal 
extraction of the spectra. The algorithm used has
been developed by Horne 
(1986), and is efficient in optimizing S/N as 
well as in cosmic rays removal. The resulting one-dimensional spectra were\\ 
b) flat-fielded \\
and \\
c) transformed to a wavelength scale using He--Ne lamp calibrations 
before and after each exposure, to correct for deflections in the 
spectrograph. Given the importance of a good wavelength calibration 
for the final result, several 
quality checks were applied, including control of the position of night-sky 
lines in galaxy spectra and self-consistency in the position of different 
emission lines where more than one was found. A precision of better than 
\mbox{\plumi 1 \AA}
everywhere in the spectrum was obtained, which translates to 70 \kms 
uncertainty per line. Afterwards, the spectra were\\
d) corrected for the characteristic transmission 
of the fibres and then \\
e) the sky was subtracted. Unfortunately, our night sky 
exposures were severely read-out noise limited. A better approximation for
the sky contribution was provided by the faintest galaxy spectrum out of 
all those taken with the same OPTOPUS-exposure.
The results were also tested with the second faintest galaxy and proved to be
reliable for spectra with a S/N above 7. This was fulfilled 
in \mbox{70 \%} of the cases. The reason for this high success
rate is that the range in brightness between the faintest and the brightest 
galaxies in the small field of view is necessarily large and thus the faintest
galaxies are strongly underexposed.\\
After these stages the spectra are finally ready for redshift measurement,
either through fitting of known emission lines, or through 
cross-correlation against a template galaxy of known 
redshift, or both.

\subsection{Emission-line redshifts}
As a next step every spectrum was inspected for 
the presence of strong 
emission lines. If more than one was found, the redshifted position of each of 
them was determined through an interactive fitting procedure. Thereby a 
Gaussian superimposed onto a linear or quadratic polynomial,
approximating the local continuum, was used. As expected, the most common 
emission lines found were 
H$_{\beta}$ and the two O III lines (4959 and 5007 \AA); in some cases also 
O II, Ne III, H$_{\gamma}$ and H$_{\delta}$. 
The largest errors of the emission line redshifts 
are due to the uncertainties in the fit as well as to 
bigger-than-average deviations in the wavelength calibration quality at
the position of a given emission line.
An error analysis of a subsample of emission 
line spectra shows that the measurement errors amount on average 
to about 100 \kms \label{emis}
for each single line. As a consequence, an uncertainty of 
100/$\sqrt{n}$ ~\kms was chosen for emission line redshifts, where $n$ stands 
for the number of emission lines per galaxy involved. A minimum of n=2 emission
lines was required to avoid misidentifications.

\subsection{Cross-correlation redshifts}
\begin{figure}[hbtp]
\centerline{\EPSFxsize=8.8cm \EPSFbox{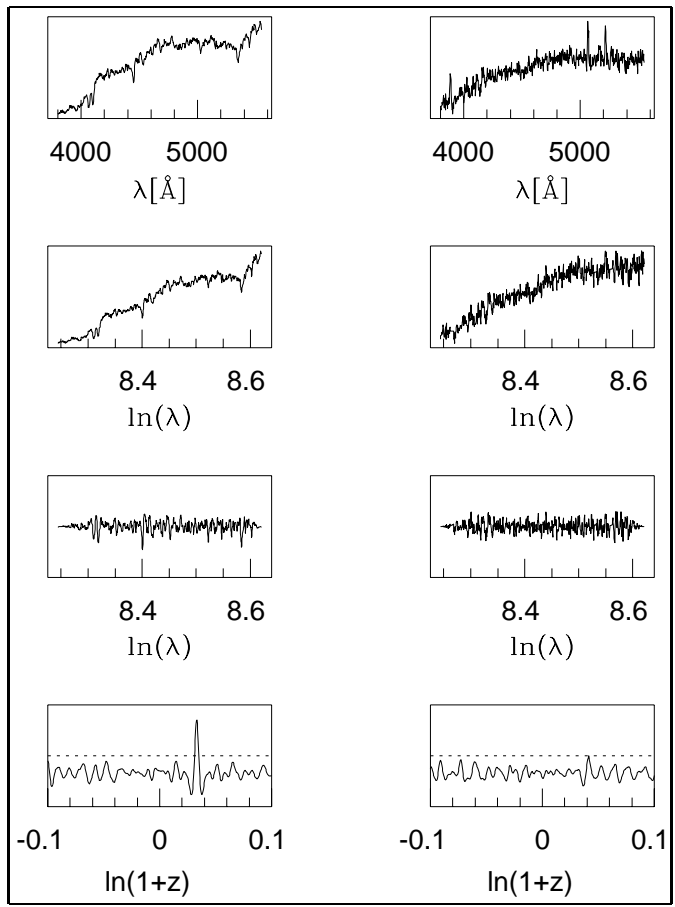}}
\caption{Cross-correlation technique. Left an example for a spectrum with 
good S/N and right a faint spectrum, which barely reaches the threshold 
for inclusion in the redshift catalogue. Strong emission lines are first
eliminated from the pre-reduced spectra (top) and the spectra are rebinned
logarithmically (2nd row). Then the continuum is subtracted (3rd row). These
spectra are cross-correlated against a template and the 
resulting function plotted (bottom). If the maximum of the normalized 
cross-correlation-function 
fails to reach the dashed line with the value of 0.25, the result is ignored.
\label{redu}}
\end{figure}

With the exception of Seyferts and other galaxies with central activity, 
no emission lines are expected in the central region of a 
galaxy covered by the OPTOPUS fibres. Most spectra are bulge-dominated and
characterized by strong absorption lines, upon which the 
cross-correlation method relies. This method was described in detail by e.g.
Tonry and Davis (1979), 
and will be abbreviated with cross-correlation-method in the following. 
In recent years cross-correlation has become a standard procedure, mainly
because of its precision and objective error analysis.
The achieved accuracy will be discussed in Sect. 5.
Prior to cross-correlation it is necessary to cut emission lines away and
transform the spectral continuum to a constant level of zero. Special 
care was taken while using a high-pass filter in order to make this 
continuum subtraction. Residual features of low spatial frequency
would result in being treated as broad, superposed ``absorption lines'', 
distorting the peak of the cross-correlation-function.\\
A fundamental problem using the cross-correlation-method is the choice of one or more 
templates, because this defines directly the zero-point shift. There are 
divergent opinions on which kind of and on how many 
different templates to use
(Teague \et 1990; Malumuth \et 1992; 
Bardelli \et 1994). For the sake of speed 
and simplicity only one template was used here, which was constructed 
merging 20 galaxy spectra with high S/N. They were 
obtained during the same observing runs as the 
data to be reduced. In this way, systematic deviations due to
instrument configuration and setup could be minimized. Moreover, we
concluded from several tests that a template with 
careful zero-point determination, reasonable 
line widths, and a scaling of all larger absorption lines to the same depth 
is more decisive for the quality of the results than using several 
different stars and galaxies as templates.\\
Some of the reduction steps after spectra preparation are illustrated in 
Fig. \ref{redu}, where a high S/N spectrum is shown, as well as one that 
barely 
reached the level for inclusion into the redshift catalogue. A threshold of 
0.25 was chosen for the height of a normalized cross-correlation-peak in order to flag 
the result as confident. This limit is meant to be conservative and was 
determined by examining the level below which cross-correlation-results became randomly
distributed instead of lying in a realistic redshift range.

\section{Catalogue}
\begin{figure}[hbtp]
\centerline{\epsfxsize=8.8cm \epsfbox{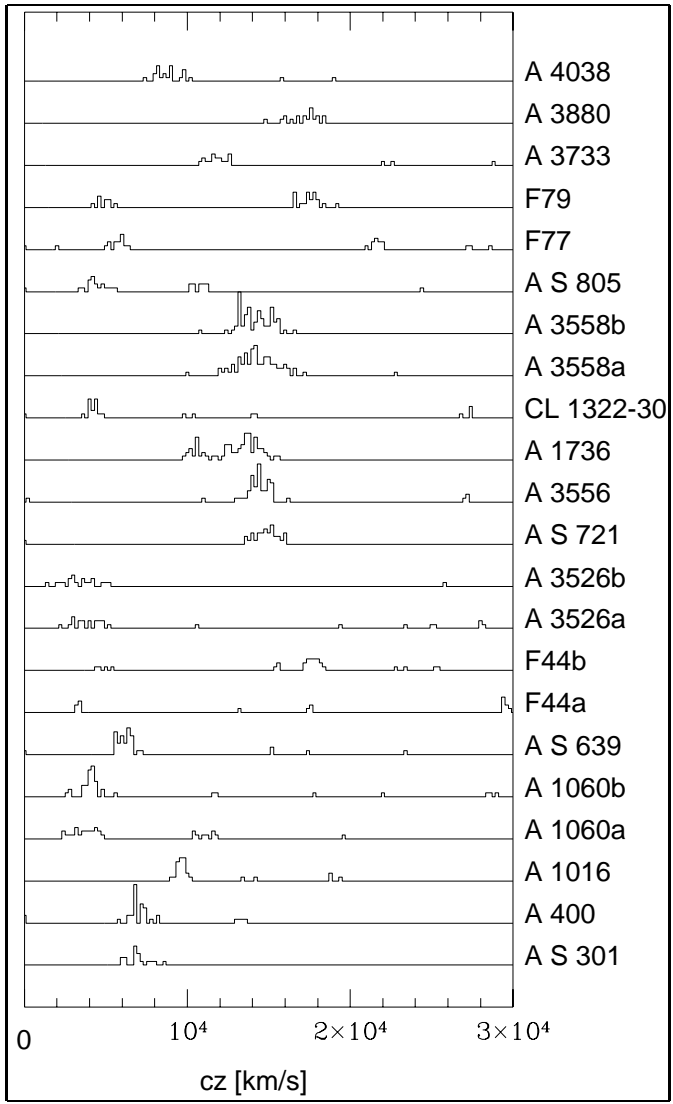}}
\caption{Histogram of redshifts measured in this work, divided into
OPTOPUS fields. In fields 
A S~301, A 400, A 3556, A 1736, A 3558a, A 3558b, A S~805, A 3733, A 3880 
and A 4038 more than 
one exposure has been taken. The contribution of fore- and background
galaxies is clearly visible.
The offset between zero-level lines 
in vertical direction amounts to 11 galaxies.
\label{histo}}
\end{figure}

In cases where both emission and absorption line redshifts could be 
measured for the same galaxy, they always agreed within statistics and
a weighted mean was taken.\\
The result of the different reduction steps and cross-correlation described 
above is a 
catalogue with 860 redshifts. The data are listed in Table 1 and 2 and 
velocity histograms are drawn for each cluster separately in Fig. \ref{histo}. 
The galaxies have been divided into only three 
morphological classes, E, S0 and S, because of the impossibility to extract
more information from direct plate examination. 
It should be stressed 
that this is not meant to be a definitive morphological classification.
The correspondence between the author's classification 
and literature data, where available, is around 80\%. This information will be
used in a forthcoming paper in order to separate statistically, using a large 
sample, kinematical effects into distinct morphological classes. 

\section{Redshift Errors} 
\label{erro}

The measured redshifts can be affected by various error sources.
For instance, poor 
flat-fielding could cause a spurious distortion of relevant lines 
and features. For faint sources good sky subtraction is of paramount 
importance. Moreover, the quality of the wavelength calibration 
depends directly on the spacing and on the S/N of comparison lines. 
Some other minor error sources 
exist, like possible intrinsic dissimilarities between the spectra of 
template and object or inaccurate fitting of the cross-correlation-peak. 
All of these effects result in a distortion of the cross-correlation-peak 
and are thus reflected in the
estimation of redshift uncertainties (see Tonry \& Davis 1979 for a detailed
description of error estimation). \\
With the cross-correlation-method only relative estimates for the random redshift errors
are possible. It is therefore necessary to apply a scaling factor to 
these errors in order to obtain realistic values. In a first step such a 
scaling factor has been determined using those 36 galaxies 
which had been observed twice (see also Malumuth \et 1992).\\
Of course, the corrected errors, listed in Table 1 and 2, may still be 
smaller than the true external random errors. However, the subsequent 
comparison with redshifts from other sources in Sect. 6 and the resulting 
consistency can be taken as evidence that the errors listed in
Tables 1 and 2 are close to the true external errors.

\section{Comparison of Different Redshift Catalogues}
\label{comp}
Given the nearness of our cluster sample, a considerable overlap
with the literature was expected to arise. Indeed, the redshift overlap
with external sources, which existed when this program was started, was much
increased due to the publication of large amounts of additional data
(Teague \et 1990,
Beers \et 1991b, Malumuth \et 1992). 
In addition, first results from running projects could be considered (Mazure 
1994), which allowed the compilation of 
a comparison sample with more than 200 galaxies. These 
overlapping data are used in the following to estimate the external 
zero-point error and the true random errors of the galaxies in Table 1
as well as of those in external catalogues.

The procedure used for testing if two samples of redshift (radial velocity)
measurements for the same galaxies were consistent within the errors goes as
follows: let 
$v_{i}$ and $v_{j}$ be the radial velocities for the same galaxy from two 
sources; the quoted errors being $\Delta v_{i}$
and $\Delta v_{j}$, respectively. 
We define $\delta_{ud}(i,j)$ as the pairwise unweighted 
differences between velocities, and $\delta_{wd}(i,j)$ the same as 
$\delta_{ud}(i,j)$, but weighted by the errors:

\begin{equation}
\delta_{ud}(i,j) = v_{i}~\min ~v_{j}\label{eq1}
\end{equation}

\begin{equation}
\delta_{wd}(i,j) =\frac{v_{i}~\min ~v_{j}}{\sqrt{\Delta v_{i}^2\plu\Delta
v_{j}^2}}\label{eq2}
\end{equation}

Instead of taking classical 
estimators like mean and standard deviation, measures of the central
location $\mu$ and scale $\sigma$ for $\delta_{ud}$ and $\delta_{wd}$ have 
been used which are less sensitive against outliers (``biweight estimators'', 
see Beers \et 1991a).\\
Central locations of $\delta_{ud}$ and $\delta_{wd}$ give a measure of the
zero-point shift between samples i and j, while their spread is caused by
the combined random errors $\Delta v_{i}$ and $\Delta v_{j}$. In case of
negligible zero-point shift and of appropriate estimates of the random errors, 
$\delta_{wd}(i,j)$ is expected to be Gaussian distributed around 0 with
a standard deviation of 1. 
The values $\delta_{wd}$ can be interpreted as factors
with which the published redshift errors should be multiplied if the
random errors in Table 1 are correct.

\tabcolsep0.42cm
\setcounter{table}{2}

\begin{table*}[hbtp]
\caption{Internal and external comparison of redshifts. 
The comparison data was divided into three distinct groups: 
group a) included sources 2, 3, and 4, the three most modern ones,
presenting results for individual clusters.
Group b) was built by sources 5, 6, and 7, which were somehow older but
also dealing with specific clusters.
In the third group (c) all sample catalogues were put together, mainly the
RC3 (de Vaucouleurs \et 1991) and several other sources from ZCAT 
(Huchra, 1991).
Column 3 gives the number of objects involved. 
Robust measures for central location and scale (Beers \et 1991a) 
of unweighted differences are listed in columns 4 and 5,
whereas the same parameters, but 
weighted by the redshift errors are listed in columns 6 and 7.
Values in column 4 and 6 are positive if the redshift from the present work 
is bigger than the comparison redshift. The quoted intervals correspond
to \mbox{68 \%} confidence levels and were computed using a bootstrapping 
technique.}
\begin{tabular}[l]{lllD{x}{\plumi}{-1}D{x}{\plumi}{-1}
D{x}{\plumi}{-1}D{x}{\plumi}{-1}lll}
\onerule
\multicolumn{1}{c}{Seq.} & 
\multicolumn{1}{c}{Comparison} & 
\multicolumn{1}{c}{} & 
\multicolumn{2}{c}{unweighted} & 
\multicolumn{2}{c}{weighted} &
\multicolumn{1}{c}{group}\\
\multicolumn{1}{c}{} & 
\multicolumn{1}{c}{source} & 
\multicolumn{1}{c}{$N_{tot}$} & 
\multicolumn{1}{c}{$\mu_{ud}$} & 
\multicolumn{1}{c}{$\sigma_{ud}$} & 
\multicolumn{1}{c}{$\mu_{wd}$} & 
\multicolumn{1}{c}{$\sigma_{wd}$}\\
\multicolumn{1}{c}{(1)} & 
\multicolumn{1}{c}{(2)} & 
\multicolumn{1}{c}{(3)} & 
\multicolumn{1}{c}{(4)} & 
\multicolumn{1}{c}{(5)} & 
\multicolumn{1}{c}{(6)} &
\multicolumn{1}{c}{(7)} &
\multicolumn{1}{c}{(8)} \\
\onerule                    
1 &cross-corr. vs. emission      &29 &     21 x  13 & 67 x  9 &    0.31 x
 0.19 &0.99 x 0.16 &\\
\onerule                    
2 &Mazure (1994)        &31 &\min  1 x  13 & 65 x  8 &    0.02 x 0.19 &0.96 
x 0.11 &\\
3 &Malumuth \et (1992)  &20 &      5 x  14 & 60 x  9 &    0.02 x 0.26 &1.04 x 
0.14 & a\\
4 &Beers \et (1991b)    &26 &     15 x  13 & 67 x 10 &    0.22 x 0.22 
&1.09 x 0.20 &\\
\onerule 
5 &Teague \et (1990)    &43 &      5 x  16 &117 x 22 &    0.02 x 0.22 &1.47 x 
0.20 &\\
6 &Dressler \& Shectman (1988)&40 &\min 45 x  12 & 81 x 11 
&\min0.96 x 0.25 &1.70 x 0.24 & b\\
7 &Lucey \et (1986)     &33 &     43 x  25 &140 x 19 &    0.44 x 0.31 
&1.84 x 0.35 &\\
\onerule                    
8 &RC3    &52 &\min  5 x  13 &101 x 13 &\min0.31 x 0.34 &2.37 x 0.29 &\\
9 &AAT \& Stromlo &43 &\min  4 x  10 & 76 x 18 &    0.04 x 0.30 &2.02 x 0.29 &\\
10 &Fairall&13 &\min 27 x  32 & 93 x 86 &\min0.09 x 0.54 &1.84 x 0.33 & c\\
11 &Las Campanas &35 &\min 38 x  15 & 86 x 14 &\min0.68 x 0.30 &1.69 x 0.25 &\\
12 &Palumbo&11 &\min127 x 101 &245 x 54 &\min0.67 x 0.49 &1.31 x 0.23 &\\
13 &ESO    &36 &\min 35 x  32 &153 x 29 &\min0.72 x 0.61 &3.40 x 0.70 &\\
\onerule
\end{tabular}          
\end{table*}

\subsection{Internal comparison}
In a first step, cross-correlation-redshifts were compared with emission line redshifts
for those 29 galaxies for which both could be determined. 
The adopted random errors of the 
emission-line redshifts were discussed in Sect. 3.1; they amount to 60
\kms on average. The error of the cross-correlation-redshifts becomes then, inverting 
Eq. (\ref{eq2}), 34 \kmx. The zero-point difference of the two sets
is 21 \plumi 13 \kms (the emission-line redshifts being smaller);
this is hardly significant.

\subsection{External comparison}
The results of the internal check, as well as
an extensive comparison between our redshifts and literature data are
presented in Table 3. The data from the redshift collection of 
J.~Huchra at the Center of Astrophysics, Harvard 
(ZCAT) were adopted with the original source given there. 
All velocities without quoted errors were given an uncertainty 
value of 100 \kmx.

Extraction of mean zero-point shift in the redshift of the present work 
was done using only catalogues of group (a), as well as from (a) and (b)
together. The mean weighted zero-point shift amounts to -3.9 \plumi\ 5.8 \kms 
using only (a), and to -6.9 \plumi\ 4.4 \kms using (a) and (b) together. This
shift is small and definitely consistent with 0. Therefore no zero--point 
correction has been applied to the redshifts.

\begin{figure}[hbtp]
\centerline{\epsfxsize=8.8cm \epsfbox{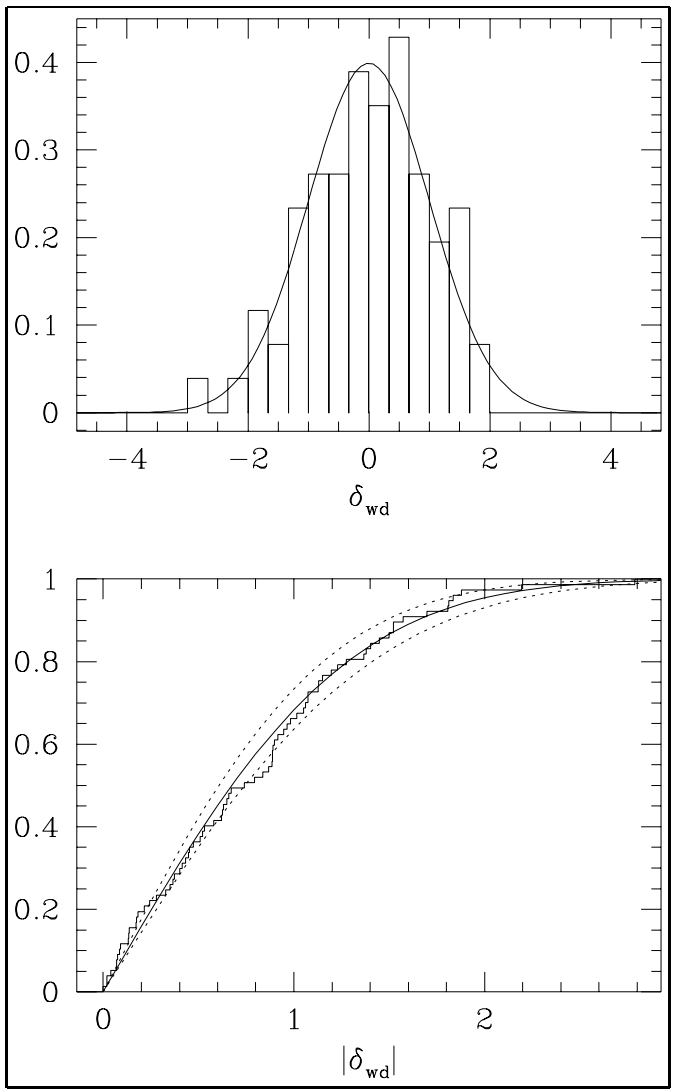}}
\caption{Top: comparison between the redshift of the present work and those 
of Mazure (1994), Malumuth \et (1992)
and Beers \et (1991b). The histogram represents the differences weighted by
the errors quoted in Table 1. 
The overlayed distribution is expected, 
under the assumption that all sources state realistic external redshift errors.
Bottom: Kolmogoroff-Smirnov test. The probability that the histogram is drawn 
from the overlayed gaussian distribution is 66\%; a $\chi^2$-test 
gave 96\% (see text). The dashed lines are cumulative gaussians with 
$\sigma = 0.9$ resp. $1.1$.
\label{vglmaz}}
\end{figure}

It can be seen that an excellent agreement exists when considering
only those recent studies which used the cross-correlation-technique and which 
concentrated on a few clusters (i. e. Mazure
1994, Malumuth \et 1992 and Beers \et 
1991b). Although the mean errors 
quoted by these three sources lie about 1.5--2 times higher than those given
in the present work no discrepancy is found (Fig. \ref{vglmaz}). 
The probability 
that $\delta_{wd}$ in this case is gaussian distributed with $\mu$ = 0 and 
$\sigma$ = 1 can be determined by a $\chi^2$-test and resulted to be 96 \%.

If we took larger redshift errors for our data a discrepancy would arise: 
with errors a factor 1.3 bigger the \mbox{$\chi^2$-test} would give
only 38 \% probability, and a factor of 1.7 would be consistent with 
$\sigma$ = 1 only with 5 \% probability. We conclude therefore that the most 
probable mean external 
error in radial velocity of our catalogue is around 30 \kmx, and in fact
smaller than 45 \kms with 95 \% confidence. An excellent
consistency between several redshift catalogues is thus found at this level 
of accuracy (compare discussion in Malumuth \et 1992). It should be stressed
that this conclusion is supported by as many as three
independent sources with a total of 77 objects involved.\\
Consequently, in view of the following considerations, we take our scaling
factor for conversion to external 
errors as confirmed and interpret any residual discrepancy as arising from
the accuracy of the external data. It can be seen in Table 3
that the only source revealing a significant zero-point shift (45 \kmx) 
relative to our data is Dressler \& Shectman (1988). 
Low values for the absolute differences $\delta_{ud}$ are 
found for Dressler \& Shectman (1988), AAT \& Stromlo and Las Campanas,
meaning measurements of high relative precision. 
Their quoted uncertainties seem to be underestimations, 
though, as can be seen from the value of $\delta_{wd}$. On the 
contrary, the catalogue of Palumbo \et lists radial velocities with modest
accuracy, but the quoted random errors seem to be realistic. Finally, 
Teague \et (1990) as
well as Fairall include data of very inhomogeneous quality in their
catalogues. In general, there are some indications for those galaxies with the 
largest errors (100--200 \kmx) to be less accurate than stated, which could 
be explained by the fact that errors of this magnitude are mostly rough
estimates.

\section{Conclusions}
New redshift measurements for 735 galaxies in the central regions 
(R = 0.2--0.8 Mpc for H$_0$ = 50 \kmx Mpc$^{-1}$) of 15 nearby 
clusters have been presented. Given a rather large overlap with existing 
data, and the homogeneity of the present measurements, it was shown that 
this redshift catalogue can be used for redshift errors calibration 
purposes, offering a
link between numerous modern catalogues. A comparison of 
only the most recent and highest-quality datasets shows that a mean external 
precision of $\la 45$ \kms has been achieved with a probability of 
\mbox{95 \%}. 
In addition, raw morphological types have been determined for most of 
the sample galaxies, using existing photographic material. These data will 
be used in a forthcoming paper in order to study substructures, as well as
kimematics of galaxy clusters in dependence of luminosity and morphological 
types.

\begin{acknowledgements}
The author thanks Drs.~A.~Schr\"oder and especially 
B.~Leibundgut for their extensive contributions in the early 
stages of this project. I am grateful to Proff.~I.~Appenzeller
and R.~Bender for their hospitality at the Landessternwarte Heidelberg and 
for many stimulating discussions. Also, I wish to thank Prof.~G.~A.~Tammann
for his support throughout the duration of this project and for careful 
reading of the manuscript. Support of the Swiss National Science 
Foundation is gratefully aknowledged.
\end{acknowledgements}


\end{document}